**Deep learning-assisted active metamaterials with heat-enhanced thermal transport**

*Peng Jin, Liujun Xu, Guoqiang Xu, Jiaxin Li, Cheng-Wei Qiu*, and Jiping Huang**

Peng Jin, Jiping Huang
Department of Physics, State Key Laboratory of Surface Physics, and Key Laboratory of Micro and Nano Photonic Structures (MOE), Fudan University, Shanghai 200438, China
E-mail: jphuang@fudan.edu.cn

Liujun Xu, Guoqiang Xu, Jiaxin Li, Cheng-Wei Qiu
Department of Electrical and Computer Engineering, National University of Singapore, Singapore 117583, Singapore
E-mail: chengwei.qiu@nus.edu.sg

Liujun Xu
Graduate School of China Academy of Engineering Physics, Beijing 100193, China

Heat management is crucial for state-of-the-art applications such as passive radiative cooling, thermally adjustable wearables, and camouflage systems. Their adaptive versions, to cater to varied requirements, lean on the potential of adaptive metamaterials. Existing efforts, however, feature with highly anisotropic parameters, narrow working-temperature ranges, and the need for manual intervention, which remain long-term and tricky obstacles for the most advanced self-adaptive metamaterials. To surmount these barriers, we introduce heat-enhanced thermal diffusion metamaterials powered by deep learning. Such active metamaterials can automatically sense ambient temperatures and swiftly, as well as continuously, adjust their thermal functions with a high degree of tunability. They maintain robust thermal performance even when external thermal fields change direction, and both simulations and experiments demonstrate exceptional results. Furthermore, we design two metadevices with on-demand adaptability, performing distinctive features with isotropic materials, wide working temperatures, and spontaneous response. This work offers a framework for the design of intelligent thermal diffusion metamaterials and can be expanded to other diffusion fields, adapting to increasingly complex and dynamic environments.

## 1. Introduction

Heat management plays a crucial role in modern technology, as efficient thermal control can substantially improve the performance and longevity of electronic devices, reduce energy consumption, and ensure the comfort and safety of users. In recent years, metamaterials-based groundbreaking innovations have emerged in the field of heat management. At a macro- scene, we observe the advancement of passive radiative cooling systems[1] that efficiently emit heat into space, thermally adjustable wearable devices[2] that regulate temperature for user comfort, and thermal camouflage platforms[3] that adapt their infrared signature to blend with the surrounding environment. At the atomic scale, nanophononic metamaterials utilize local resonances, which leads to reduction in the thermal conductivity.[4-7] In this context, the development of advanced thermal control techniques utilizing metamaterials has gained increasing importance in recent years.[8-14] Traditional metamaterials primarily focused on static cases,[15-24] lacking tunability for variable situations. To tackle this issue, tunable metamaterials with dynamic features have emerged, covering optics,[25] acoustics,[26] mechanics,[27-28] and thermotics.[29-35] For example, advanced thermal functions have been realized, such as macroscopic thermal diodes,[29] tunable analog thermal materials,[36] path-dependent thermal devices,[37] and liquid-solid hybrid metamaterials.[38] Additionally, adaptive thermal devices are presented to maintain functionality or stability depending on application scenes.[30,39-44] Despite these advancements, the achievement of state-of-the-art self-adaptive thermal diffusion metamaterials still faces three longstanding and strong barriers. Firstly, adaptive devices with robust functions usually require extremely anisotropic parameters,[39-41,45] which are challenging to prepare from natural bulk materials. Secondly, existing adaptive devices, especially macroscopic thermal diodes[29] and energy-free temperature trapping,[30] are limited to a specific temperature range related to the phase-change temperature of shape memory alloys. Finally, most tunable or adaptive metamaterials[25-26,29-31,36-44,46] need to be adjusted manually rather than automatically, lacking self-cognitive ability.

Recently, the emergence of ChatGPT[47] is a testament to the great potential of artificial intelligence. Also, intelligent materials, involving interdisciplinary research and combining intelligent algorithms with material design, have promising applications in optics,[48] nanotechnology,[49] theoretical physics,[50] materials science,[51] and thermal science.[52] These advances have inspired the development of ideal self-adaptive thermal diffusion metamaterials that fully embrace intelligence. Ideally, these metamaterials should automatically (without human aid) and timely adjust their dynamic components to keep function stable or switch functions continuously in response to the broad range of ambient temperature change. Such self-adaptive (or say, active) thermal diffusion metamaterials would be highly desirable in situ scenes. However, the demanding technical performance requires an appropriate actuation mechanism that integrates an algorithm-driven intelligent system with thermal diffusion metamaterial design. Although these metamaterials have been used to design advanced self-adaptive optical cloaks in wave systems,[53] they have failed in diffusion systems like heat transfer due to the lack of controllable degrees of freedom. Existing machine learning-based thermal diffusion metamaterials are dictated by the inverse design methods,[54-57] which calculate the parameters of materials and sizes for desired functions. In addition, once these metamaterials are prepared, their functions are not switchable, lacking the ability to adapt to various scenes.

Here, we introduce a deep learning-assisted intelligent system into conventional metamaterial design. As a conceptual implementation, we propose heat-enhanced thermal diffusion metamaterials driven by big data. We load the pre-trained artificial neural network into a hardware system and combine it with the bilayer structure.[58] Depending on the sensing-feedback ambient temperatures, the thermal conductivity of a rotating component could be adjusted within a vast range, which leads to a large-tunable temperature gradient in the target region. These results are verified by finite-element simulations and experiments. We then present two applications with on-demand adaptability. The first application is a thermal signal modulator with functional robustness, which enhances the clarity of original thermal signals. The second application is an intelligent thermoelectric generator with adaptive functional choice, which can automatically adjust the electromotive force generated by thermoelectric materials[59] in response to ambient temperature changes. Our heat-enhanced thermal diffusion metamaterials feature isotropic materials, unlimited working temperatures, and cognitive responsiveness, see **Figure 1**. Furthermore, we develop a handy actuation mechanism that integrates deep learning-driven intelligent systems with thermal diffusion metamaterials design. Our work represents a significant advancement in the design of self-adaptive thermal diffusion metamaterials that can operate without human intervention.

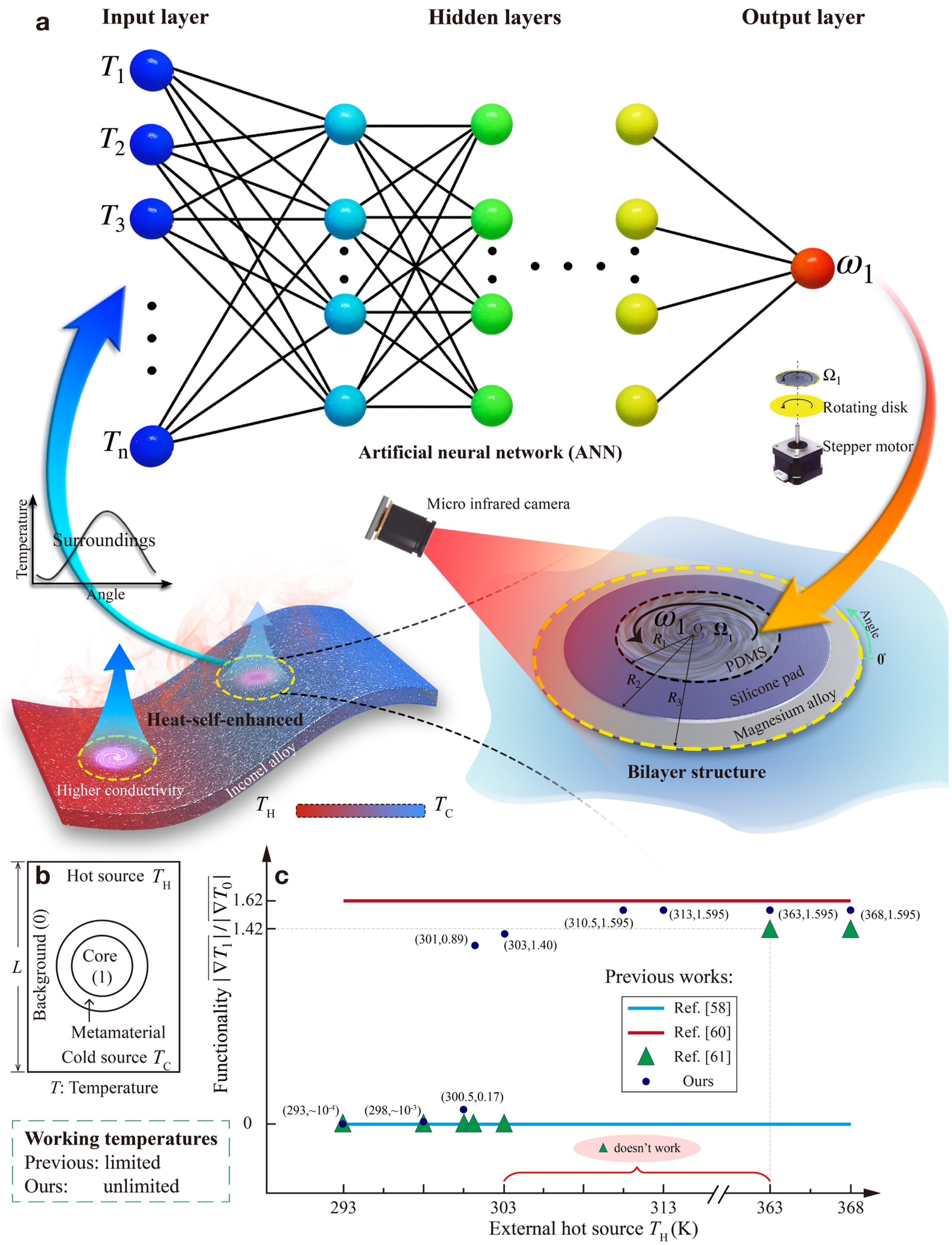

**Figure 1.** Architecture. a) Schematic of a deep learning-assisted active metamaterial. It comprises a micro infrared camera, a pre-trained artificial neural network, a stepper motor, and a bilayer structure. The infrared camera measures the temperature data ($T_1$, $T_2$, $T_3$, ..., $T_n$) of the bilayer structure's surroundings. Here, the bilayer structure consists of the inner layer (Silicone pad) and the outer layer (Magnesium alloy). The measured temperature data is input into the pre-trained artificial neural network, calculating the rotating velocity $\omega_1$ of the rotating component (PDMS) in the core region $\Omega_1$. The stepper motor reads the rotating velocity and drives the rotating component to rotate (through the rotating disk). Finally, thermal functions in the core region are regulated. b) Schematic of a common thermal metadevice. Shell-like metamaterial envelops a core region (labeled by 1), put in the background (labeled by 0). When external temperature settings are given (line source: hot source $T_\mathrm{H}$ is variable; cold source $T_\mathrm{C} = 283\ \mathrm{K}$, whose interval is $L$), metamaterial modulates the core region's temperature fields without disturbing background's counterparts. c) Comparison of functionality between finite-element simulation results from previous works[58,60-61] and ours. Here, device functions are characterized by the ratio of the temperature gradient in the core region to the external counterpart ($|\nabla T_1|/|\nabla T_0|$). The reason is that such quantity can reflect the temperature field modulation effect of the metamaterial on the external temperature field ($|\nabla T_1|/|\nabla T_0| < 1$ for cloaking; $|\nabla T_1|/|\nabla T_0| > 1$ for concentration). The ratio from some works is constant. To understand this, we give an intuitive explanation of its physical mechanism in Section S1, Supporting Information. In previous works, $|\nabla T_1|/|\nabla T_0|$ is stationary or discretely changeable within a limited working-temperature range. Ours is self-adaptive and tunable, working within an unlimited temperature range.

## 2. Architecture

The architecture of the heat-enhanced thermal diffusion metamaterial is presented in **Figure 1**a. It comprises four main modules: a temperature acquisition module (micro infrared camera), a computing system with a pre-trained artificial neural network (ANN), a stepper motor, and a bilayer structure. We aim to adjust the effective thermal conductivity of the target region based on temperature information feedback from its surroundings. Here, as a proof-of-concept implementation, we consider a two-dimensional system with a bilayer structure. The target region is the core region $\Omega_1$ consisting of poly-dimethylsiloxane (PDMS). The component of the inner layer (Silicone pad) is approximately adiabatic for precise control of thermal fields of $\Omega_1$, and the outer layer as a compensation layer (Magnesium alloy) is intended not to disturb the thermal fields of the background (Inconel alloy). The thermal conductivity from the inside out is 0.15, 1, 72.7, and 9.8 W m$^{-1}$ K$^{-1}$, respectively. After setting $R_1$ = 30 mm and $R_2$ = 53 mm, $R_3$ can be calculated as 60 mm.[58]

To characterize the temperature information of the bilayer structure's surroundings, we select several discrete positions around the outer layer and measure their temperatures using a micro infrared camera. As illustrated in Figure 1a, the yellow dashed circle with a radius of $R_3$ represents the selected bilayer structure's surroundings, and the position marked 0° denotes the first position. Next, we record the temperature values ($T_1$, $T_2$, $T_3$, ..., $T_N$) of N equally spaced positions along the circumference in a counterclockwise direction, serving as the input data of ANN [denoted by $\boldsymbol{T}^{(0)}$].

Thanks to the tunable analog thermal material,[17,36,62] by rotating (rotating velocity: $\omega_1$) the PDMS in the core region $\Omega_1$, the effective thermal conductivity of the rotating medium can be tuned from near-zero ($\omega_1$ = 0) to near-infinity (larger $\omega_1$). Here comes an intuitive picture to understand this mechanism. The effective thermal conductivity of the rotating medium is derived from the heat flow conservation equation. By incorporating the rotating convective components, such highly tunable convective heat can be equivalently converted to the highly tunable effective thermal conductivity of this medium under a limited temperature-gradient range. On the other hand, the effective thermal conductivity of $\Omega_1$ is the crucial factor affecting the temperature-gradient distributions of $\Omega_1$. Therefore, the core region's temperature-gradient distribution reflects its thermal conductivity.

To incorporate "intelligence" into the system, we utilize an ANN to establish a relationship between the extracted temperature information [input data: $\boldsymbol{T}^{(0)}$] and the rotating velocity of $\Omega_1$ (output data: $\omega_1$). In Figure 1, we show the structural component of the ANN, which is fully connected with four hidden layers (50 neurons per layer). The activations of all neurons in the next layer are determined by the activations of those in the current layer, $\boldsymbol{H}^{(i)}$, represented by

$$\begin{cases} \boldsymbol{H}^{(i+1)} = \mathrm{ReLU}\left[\boldsymbol{W}^{(i+1)}\boldsymbol{T}^{(i)} + \boldsymbol{b}^{(i+1)}\right], i = 0 \\ \boldsymbol{H}^{(i+1)} = \mathrm{ReLU}\left[\boldsymbol{W}^{(i+1)}\boldsymbol{H}^{(i)} + \boldsymbol{b}^{(i+1)}\right], 0 < i < 4 \\ \mathrm{ReLU}(\omega_1) = \mathrm{ReLU}\left[\boldsymbol{W}^{(i+1)}\boldsymbol{H}^{(i)} + \boldsymbol{b}^{(i+1)}\right], i = 4 \end{cases} \quad [1]$$

where ReLU (a) $= \max(0, \mathrm{a})$ is the rectified linear unit function. $\boldsymbol{W}$ and $\boldsymbol{b}$ are weights and biases for the neurons. $i$ is the ordinal number of layers, and $i = 0$ represents the input layer.

As the ANN is a data-driven model, we prepare the dataset (Section S2 and S3, Supporting Information). Finally, we train the proposed ANN with selected hyperparameters using the backpropagation algorithm to achieve optimal performance (Section S2, Supporting Information).

Once the thermal field of the bilayer structure reaches equilibrium, the temperature information ($T_1$, $T_2$, $T_3$, ..., $T_N$) is collected by a micro infrared camera at a circle with a radius of $R_3$ = 60 mm. This array of temperature values is then input into the computing system to obtain the corresponding output signal, which is the rotating velocity of the stepper motor, denoted as $\omega_1$. Considering a case where $\omega_1$ is 0 rad s$^{-1}$, static PDMS possesses a thermal conductivity of 0.15 W m$^{-1}$ K$^{-1}$, resulting in the maximum temperature gradient in $\Omega_1$.

The middle left inset of Figure 1a represents a schematic of a potential intelligent heat dissipation material arranged by the heat-enhanced metamaterials, and color denotes temperature distributions within a range from the lowest $T_{\mathrm{low}}$ to the highest $T_{\mathrm{high}}$ temperature. More uniform surrounding temperatures (hot end) result in higher thermal conductivity, determined by the pre-trained ANN. The principle is that a smaller (larger) external temperature gradient brings about larger (smaller) rotating velocities of the core region, resulting in a smaller (larger) temperature gradient in the core region. When the bilayer structure is under a smaller external temperature gradient (which means more uniform surrounding temperatures), it brings about larger rotating velocities, leading to higher effective thermal conductivity.

**3. Omnidirectional response in simulations**

Finite-element simulations are first employed to evaluate the performance of the heat-enhanced thermal diffusion metamaterial. For proof-of-concept verification, we consider a two-dimensional bilayer structure whose components and sizes are the same as described earlier. The system's left (right) end connects to a hot (cold) source. In our simulations, the cold source ($T_{\mathrm{C}}$ = 283 K) is fixed, while the hot source ($T_{\mathrm{H}}$) is varied. We first collect the temperature data [$T_a{}^{(0)}$, $T_b{}^{(0)}$, $T_c{}^{(0)}$] of N = 36 equally spaced positions in the yellow dashed circle in three cases ($T_{\mathrm{H}}$ = 293, 303, 313 K) of a static bilayer structure, as shown in **Figure 2**a. Here, "static bilayer structure" means that the core region is not rotating as an initial state of the active metamaterial. For each case, the first data is the temperature of the position marked 0° in the dashed circle, and the temperature of these positions is collected in counterclockwise order, serving as the input layer of the pre-trained ANN. Hence, via the pre-trained ANN, the rotating velocity $\omega_1$ of PDMS is calculated as 0.10, 0.00067, and 0 rad s$^{-1}$ for the three temperature settings, respectively. After setting the above parameters ($T_{\mathrm{H}}$ and $\omega_1$) in finite-element simulations, we show these three temperature profiles (color distributions) of the bilayer structure; see Figure 2b. No matter how the rotating velocity $\omega_1$ of PDMS changes, the temperature distributions of the background remain unaffected (Section S4, Supporting Information). Finally, the corresponding temperature-gradient distributions in $\Omega_1$ are presented in the right part of Figure 2c. For comparison, we show the temperature-gradient distributions in $\Omega_1$ of pure background (size: 200 × 200 mm$^2$) with three hot sources; see left part of Figure 2c. As anticipated, there is a mapping relationship between the lowest/highest $T_{\mathrm{H}} - T_{\mathrm{C}}$ and the highest/lowest rotating velocity $\omega_1$ (or, equivalently, the highest/lowest effective thermal conductivity in $\Omega_1$) in the above scheme. Furthermore, the range of the originally external temperature-gradient field $[(T_{\mathrm{H}}-T_{\mathrm{C}})/L$ : 50 to 150 K m$^{-1}$] can be adjusted to a wider range of temperature gradient ($|\nabla T|$: 0 to 238.5 K m$^{-1}$) in the target region $\Omega_1$.

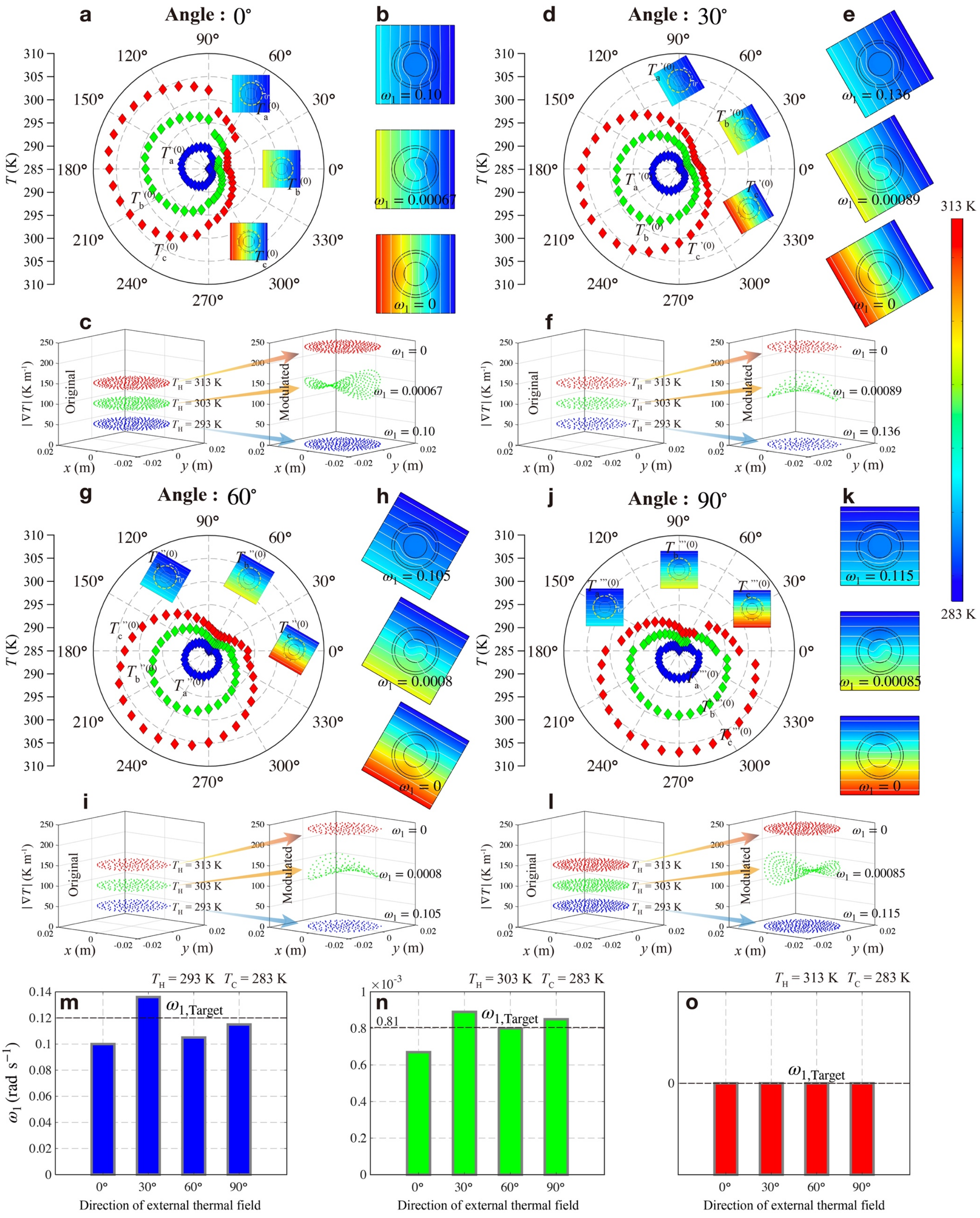

**Figure 2.** Response of the heat-enhanced thermal diffusion metamaterial in simulations. a) N = 36 equally spaced temperature data $T_a{}^{(0)}$, $T_b{}^{(0)}$, $T_c{}^{(0)}$ in the yellow dashed circle in static bilayer structures. The hot source is set as 293, 303, 313 K, respectively. The cold source is fixed at 283 K. The first data is the temperature of the position marked 0°. Each temperature

data is taken every 10° in the counterclockwise direction. b) Temperature profiles of the bilayer structure with rotating velocity $\omega_1 = 0.10$, 0.00067, 0 rad s$^{-1}$, respectively. Left part of c) Temperature-gradient distributions $|\nabla T|$ in core region $\Omega_1$ of static pure background with three hot sources. Right part of c) Temperature-gradient distributions $|\nabla T|$ in core region $\Omega_1$ of the bilayer structure with $\omega_1 = 0.10$, 0.00067, 0 rad s$^{-1}$, respectively. d-f) Same characterization with (a-c) when the external thermal field rotates 30° around the center of the bilayer structure in the counterclockwise direction. g-i) Same characterization with (a-c) when the external thermal field rotates 60° around the center of the bilayer structure in the counterclockwise direction. j-l) Same characterization with (a-c) when the external thermal field rotates 90° around the center of the bilayer structure in the counterclockwise direction. m-o) Comparison of calculated $\omega_1$ and targeted $\omega_{1,\mathrm{Target}}$ in above four directions of the external thermal field in cases with $T_\mathrm{H}$ = 293, 303, 313 K, respectively.

In the actual scene, we are unsure in which direction the external thermal field is exerted on the bilayer structure. Therefore, we should ensure that the above thermal performance remains unaffected by changes in the external thermal field's direction. In particular, we guarantee the hot and cold sources consistent with the above and rotate them 30, 60, and 90° around the center of the bilayer structure. Apparently, temperature distributions in the yellow dashed circle is different from each other when rotating the same $T_\mathrm{H}$ and $T_\mathrm{C}$; see $\boldsymbol{T}^{(0)}$, $\boldsymbol{T}'^{(0)}$, $\boldsymbol{T}''^{(0)}$, and $\boldsymbol{T}'''^{(0)}$ in Figure 2a,d,g,j. Subsequently, we calculate the rotating velocity $\omega_1$ in cases with $T_\mathrm{H}$ = 293, 303, 313 K in the above four directions using the pre-trained ANN and perform finite-element simulations. Finally, Figure 2c,f,i,l confirm that the heat-enhanced metamaterial has good robustness to changes in the external thermal flow's direction. Such intelligent metamaterial also maintains functional stability under different cold sources and non-uniform external thermal fields (Section S5, Supporting Information). In addition, the calculated $\omega_1$ from pre-trained ANN is almost consistent with the set $\omega_{1,\mathrm{Target}}$ when external hot and cold sources are given, as shown in Figure 2m-o. In Section S6, Supporting Information, we discuss the weak direction-dependence of the metamaterial's performance (reflected by calculated $\omega_1$) even though it comprises isotropic materials with isotropic geometry.

**4. Experimental measurements**

To evaluate the performance, we initially set the temperatures of the hot and cold baths to 293 K and 283 K, respectively. After starting the heat-enhanced metamaterial, it measures the temperature data $T_a^{(0)}$, and using a pre-trained ANN, it calculates the rotating velocity $\omega_1$ of PDMS, which is found to be 0.118 rad s$^{-1}$ (refer to **Figure 3**a). Further, Figure 3b shows the temperature profile of the bilayer structure recorded by an infrared camera Fotric 430. Note that in the core region, the temperature distribution is uniform, and in the background region, the temperature field is nearly undistorted. Subsequently, we fix the temperature of the cold bath to 283 K and change the temperature of the hot bath to 303 K. The measured temperature data $T_b^{(0)}$ and the calculated $\omega_1 = 0.0007$ rad s$^{-1}$ are shown in Figure 3c. The temperature profile of the bilayer structure is presented in Figure 3d, where the uniformity of the temperature distribution in the core region is slightly broken while the background temperature field is still undisturbed. Finally, with the same cold bath, we increase the temperature of the hot bath to 313 K, and read the temperature data $T_c^{(0)}$, see Figure 3e. As anticipated, the calculated $\omega_1$ is 0 rad s$^{-1}$. We then obtain the temperature profile of the bilayer structure (Figure 3f), where we observe that the temperature distribution in the core region is maximally non-uniform, without significantly disturbing the background thermal field. Above temperature data $T_a^{(0)}$, $T_b^{(0)}$, and $T_c^{(0)}$, the relevant rotating velocity $\omega_1$, and their temperature profiles of the bilayer structure are consistent with the simulation results. For quantitative analysis, we use the central difference method to process the discrete temperature data in Figure 3b,d,f and obtain these temperature-gradient distributions in the core region (see Figure 3g), which is in good agreement with the simulation results in the right part of Figure 2c. As a result, we successfully manipulate the core region's temperature gradient in the bilayer structure (or, equivalently, the effective thermal conductivity in $\Omega_1$), based on the feedback of temperature information $T^{(0)}$ from its surroundings.

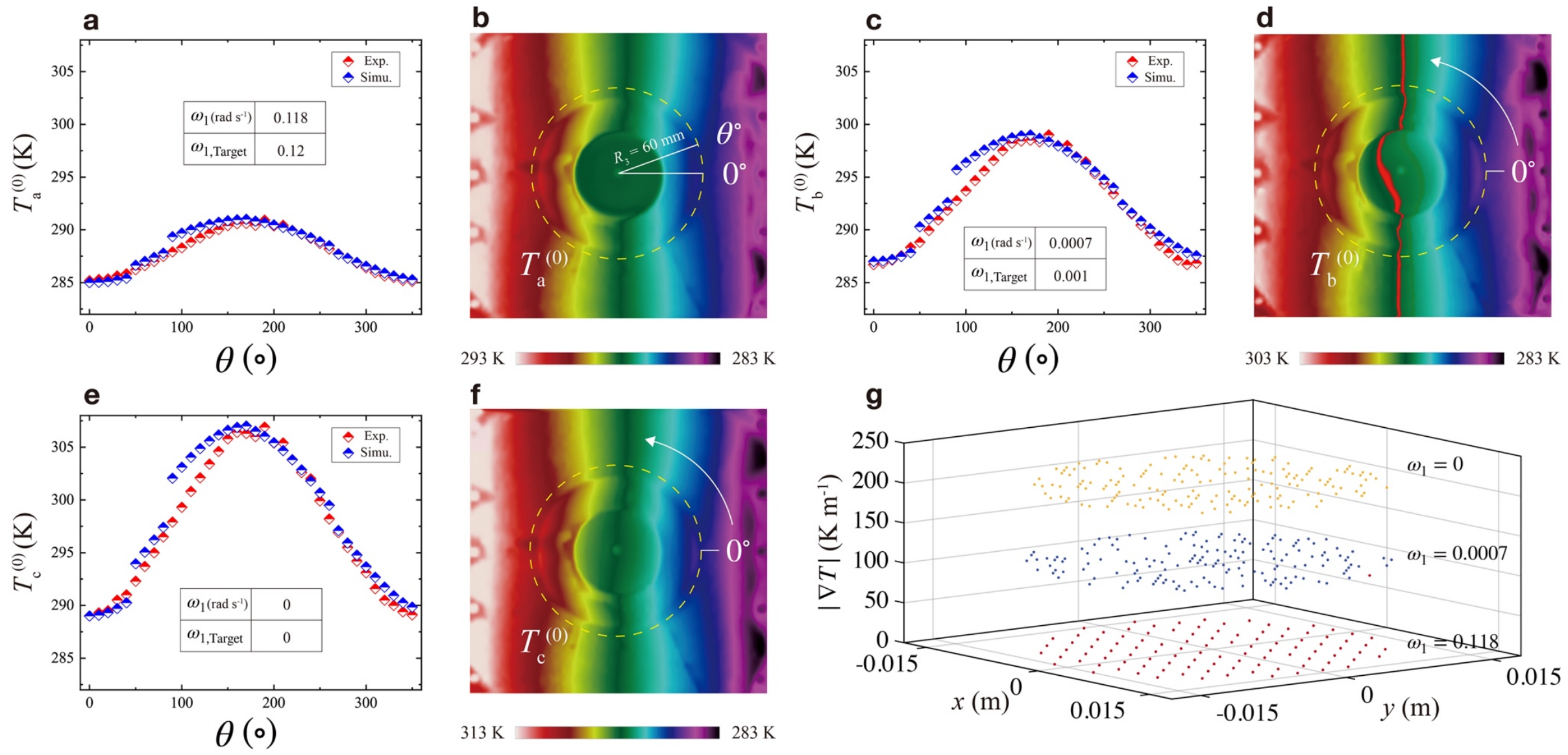

**Figure 3.** Realization of the heat-enhanced thermal diffusion metamaterial. a,c,e) Experimental temperature data ${T_{\boldsymbol{a}}}^{(0)}$, ${T_{\boldsymbol{b}}}^{(0)}$, ${T_{\boldsymbol{c}}}^{(0)}$ in the dashed circle with radius $R_3$ = 60 mm, marked in (b,d,f) when hot bath is set to 293, 303, 313 K, respectively. The cold bath is fixed to 283 K. b,d,f) Measured temperature profile of the bilayer structure with rotating velocity $\omega_1 = 0.118$, 0.0007, 0 rad s$^{-1}$, respectively. g) Calculated temperature-gradient distributions $|\nabla T|$ in core region $\Omega_1$ of the bilayer structure with $\omega_1 = 0.118$, 0.0007, 0 rad s$^{-1}$, respectively.

## 5. Heat-enhanced thermal signal modulator

We have developed a thermal signal modulator that utilizes the heat-enhanced thermal diffusion metamaterial for heat communications. Previous work adopted binary thermal spatial coding to store information in thermal signals.[63] Binary 0 and 1 are represented by encoding the temperature gradient in the working zone of the cloaking and concentration with core-shell structures, where the core region is the working zone. For thermal cloaking (concentration), there is a minimum (maximum) temperature-gradient distribution in the working zone. Via the continuous arrangement of cloaking or concentration devices, thermal signals could be stored and characterized by the temperature-gradient distributions in the working zones of the arranged metadevices. However, binary encoding is inefficient for information storage. Instead, thermal signals should oscillate with space in a continuous mode to transmit more encoding information simultaneously in heat communications. Nevertheless, the adoption of continuous encoding can result in original thermal signals being easily disturbed, with oscillations limited to smaller amplitudes due to thermal dissipation and thermal noise. Our proposed thermal signal modulator can re-modulate original disturbed thermal signals to oscillate with space in a larger amplitude, see the schematic in **Figure 4**a. To create an encoding zone, we select a square area (see the dashed box marked in Figure 4a) in the working zone of devices that is consistent with the size of the heat-enhanced thermal diffusion metamaterial. To ensure the temperature field in the working zone of the device is not disturbed, we make the effective thermal conductivity inside and outside the encoding zone consistent. We consider the original temperature gradient in the encoding zone as the external thermal field of the heat-enhanced metamaterial. In our simulations, external thermal fields $|\nabla T|$ vary from $(T_{\mathrm{H,min}} - T_{\mathrm{C}})/L$ = 50 to $(T_{\mathrm{H,max}} - T_{\mathrm{C}})/L$ = 150 K m$^{-1}$. Here, $T_{\mathrm{H,max}}$ ($T_{\mathrm{H,min}}$) represents the highest (lowest) temperature in encoding zone. $L$ is the side length of the encoding zone. Through the heat-enhanced thermal diffusion metamaterial, the average temperature gradient of the modulated zone ($\Omega_1$; see round dashed line marked in Figure 4a) could be ranged from 0 to 238.5 K m$^{-1}$. We obtain a linear transformation relationship between the original temperature-gradient range and modulated temperature-gradient range.

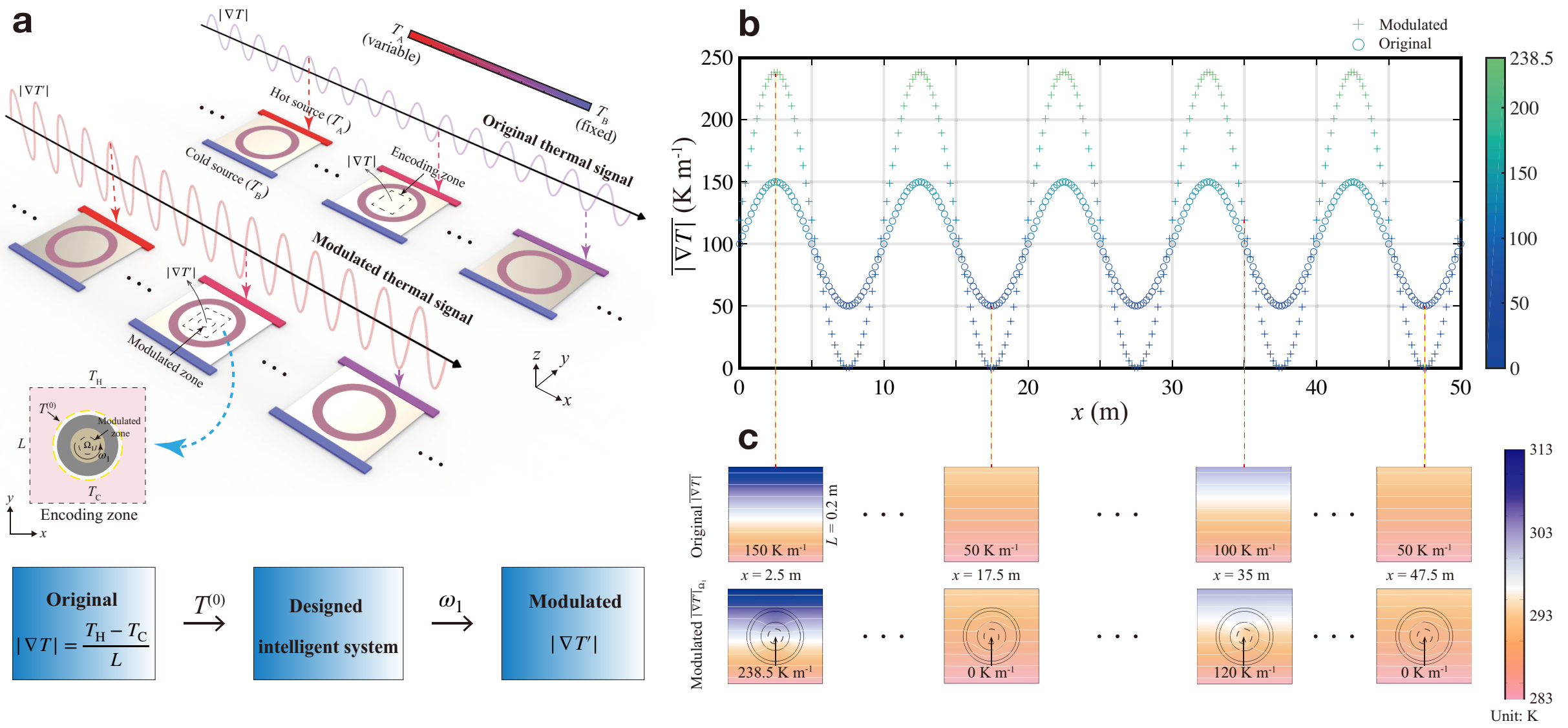

**Figure 4.** Heat-enhanced thermal diffusion metamaterial for robust modulation of thermal signals. a) Schematic of the thermal signal modulator. b) Comparison of modulated and original thermal signals. Modulated (original) thermal signals are denoted by distributions of averaged temperature gradients in the modulated zones of the modulators (only external thermal fields) across the $x$ direction. Each $x$ coordinate represents the central position of each modulator. c) Simulated temperature profiles of several modulators and their external thermal fields placed at $x$ = 2.5, 17.5, 35, and 47.5 m, respectively.

We consider the oscillation of original thermal signals driven by variable hot sources across the $x$ direction in the sinusoidal law $\sin 2\pi x/\lambda$, and apply the linear transformation to the original thermal signals to get the modulated thermal signals, see Figure 4b. We take $\lambda$ as 10 m, and arrange 50 modulators (each size: 0.2 × 0.2 m$^2$) in each range of $\lambda$. Here, $\lambda$ is the wavelength of sinusoidal thermal signals oscillating with space. Figure 4c shows the temperature profiles of several modulators and their original external thermal fields placed at $x$ = 2.5, 17.5, 35, and 47.5 m (from finite-element simulations), respectively. Here, each $x$ coordinate represents the central position of each modulator. We mark the original (modulated) temperature gradient values in the encoding (modulated) zone of several modulators; see Figure 4c. Overall, heat-enhanced thermal signal modulator preserves the relative strength relationship between the original signals while making the difference between the original signals more obvious. In Section S7, Supporting Information, we provide the mechanism of how one thermal signal modulator works.

**6. Thermoelectric generator**

In our experiment, we adjust the rotating velocity $\omega_1$ of the PDMS in the core region based on the feedback of temperature information $\boldsymbol{T}^{(0)}$ of the bilayer structure's outer layer, as shown in **Figure 5**a. This leads to a large temperature-gradient tunability in the bilayer structure's core region. Here, by the grace of these results, we propose a potential application of a heat-enhanced thermal diffusion metamaterial for thermoelectric generation (or, say, an intelligent thermoelectric generator), whose schematic diagram is shown in Figure 5b. We perform finite-element simulations to demonstrate the properties of the intelligent thermoelectric generator. In the following simulations, the components, sizes, and boundary conditions of the bilayer structure are set to be consistent with those in the experiment. Then, we place a cylindrical thermoelectric material $Bi_2Te_3$ (thermal and electrical conductivity $\kappa = 1.6$ W m$^{-1}$ K$^{-1}$ and $\sigma = 8700$ S m$^{-1}$, Seebeck coefficient $S$ = 0.2 mV K$^{-1}$) with a diameter and thickness of 30 mm and 2 mm in the center of the PDMS. $Bi_2Te_3$ contacts with the upper surface of the PDMS. Due to the thermal conductivity mismatch between $Bi_2Te_3$ and PDMS, the existence of $Bi_2Te_3$ disturbs the original temperature distributions in the core region of the bilayer structure. To mitigate this issue, we punch a series of hollow air holes with a diameter of 2 mm on $Bi_2Te_3$ to make the thermal conductivity of the two materials closer. The cold source is fixed at 283 K, and the hot source is set at 313, 303, and 293 K, respectively. The corresponding $\omega_1$ values are 0, 0.0007, and 0.118 rad s$^{-1}$, with the same conditions in the experiment. Figure 5c,f,i show the temperature profiles of the intelligent thermoelectric generator under the three simulation conditions. Note that the isotherms on the surface of $Bi_2Te_3$ become sparser and sparser with an increase in $\omega_1$, which indicates the existence of $Bi_2Te_3$ has little influence on the original temperature gradient in the core region. Next, we simulate the electric potential profiles of $Bi_2Te_3$ under the three temperature distributions in Figure 5c,f,i, as shown in Figure 5d,g,j. See concrete electric potential data of Figure 5d,g,j in Figure 5e,h,k, and the relevant maximum potential difference $\Delta U = U_{\max} - U_{\min}$ is 4.03 × 10$^{-4}$, 2.58 × 10$^{-4}$, and 2.05 × 10$^{-9}$ V, respectively. Finally, we realize the intelligent thermoelectric generator to adjust the output electromotive force adaptively under different temperature environments. For example, a relatively high ambient temperature difference makes the $\omega_1$ of PDMS in the core region zero. In this case, the temperature gradient in the core region is the largest. With the aid of $Bi_2Te_3$, the largest electromotive force difference could be generated. Interestingly, the electromotive force could drive an external cooling load, and its tunability leads to regulation of the cooling power of the load, achieving the effect of feedback regulation of the ambient temperature.

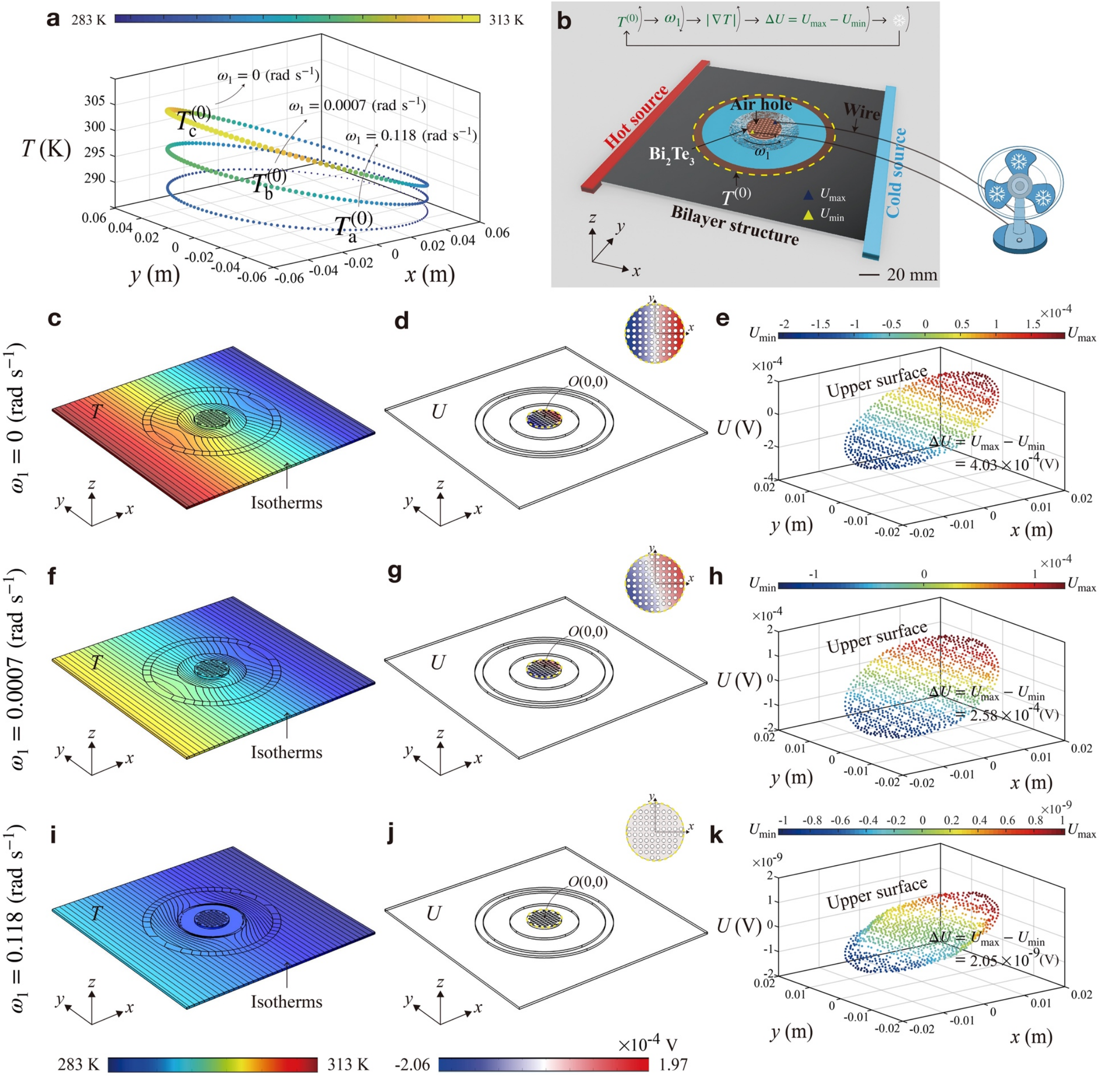

**Figure 5.** Heat-enhanced thermal diffusion metamaterial for a thermoelectric generator. a) Mapping relationship between input data $\boldsymbol{T}^{(0)}$ and output data $\omega_1$ of the intelligent metadevice from the experimental results. b) Schematic of the thermoelectric generator. c,f,i) Temperature profiles (color) of the thermoelectric generator with PDMS's rotating velocity $\omega_1 = 0$, 0.0007, 0.118 rad s$^{-1}$, respectively. Black lines represent isotherms. d,g,j) Electric potential profiles (color) of the thermoelectric generator with $\omega_1 = 0$, 0.0007, 0.118 rad s$^{-1}$, respectively. e,h,k) Electric potential distributions of the upper surface of the thermoelectric generator with $\omega_1 = 0$, 0.0007, 0.118 rad s$^{-1}$, respectively.

## 7. Conclusion

To sum up, we propose heat-enhanced thermal diffusion metamaterials, driven by a deep-learning algorithm that requires no manual intervention. As a conceptual verification, we demonstrate the intelligent adjustment of the effective thermal conductivity of the rotating component [made of poly-dimethylsiloxane (PDMS)] of a bilayer structure, based on the ambient temperature using a pre-trained artificial neural network (ANN). For algorithm implementation, we use the temperatures at discrete positions on the outer contour of the bilayer structure (ambient temperature) as input data. Then, the pre-trained ANN outputs the PDMS's rotating velocity ( $\omega_1$ ), which adjusts its effective thermal conductivity and, thus, the temperature-gradient distribution in the core region. For hardware implementation, we integrate the pre-trained ANN into a Raspberry Pi microcomputer. One end is connected to a micro infrared camera to detect the ambient temperature for the input data, and the other is connected to a stepper-motor driver to control the $\omega_1$ of the motor, rotating the PDMS. Finite-element simulations and experiments confirm the effectiveness of this heat-enhanced thermal diffusion metamaterial, which is robust to the direction of external thermal fields. In Section S8 (S9), Supporting Information, we discuss the effect of shape (length scale) on the metamaterial's performance.

Incidentally, self-adaptive or active devices can be applied in two typical scenarios. The first involves a device that maintains stable function in a changeable environment, like the thermal signal modulator we designed. The second scenario involves a device that can intelligently choose its function based on changes in its environment. To demonstrate this, we designed an intelligent thermoelectric generator that can adjust its function in response to a changeable environment. The adjustable temperature gradient in the core region enables thermoelectric materials like $Bi_2Te_3$ to generate tunable electromotive force, as verified by finite-element simulations. Finally, we have shown that metamaterials can have the ability to perceive their environment deeply. Not only that, drawing parallels from nonlinear optics,[64] our work results in material's effective thermal conductivity being responsive to external temperature gradients, laying the foundation for a configurable nonlinear thermal material. The proposed concept combines diverse domains, such as artificial intelligence, metamaterials, energy utilization, heat communications, and thermal management. The presented interdisciplinary work promises to provide new inspiration for progress in various areas, for example, intelligent thermal management in chips.

## 8. Experimental Section

The heat-enhanced metamaterial comprises four main parts: a micro infrared camera, a computing system equipped with a pre-trained artificial neural network (ANN), a stepper motor, and a bilayer structure; see the experimental setup in Figure S15, Supporting Information. The bilayer structure, with a 2 mm thickness, is connected to hot and cold containers on either side, serving as heat baths. Its components and sizes are the same as those used in the simulations. The micro infrared camera is controlled by the computing system. Every time the infrared camera is started, it measures the temperature distribution of the bilayer structure and transmits the temperature data to the computing system. The computing system consists of power, a microcomputer Raspberry Pi, a power supply of motor driver, and a motor driver. A pre-trained ANN program runs in the Raspberry Pi. The input data $\boldsymbol{T}^{(0)}$ is from the temperature data measured by the micro infrared camera. After the program processing, the computing system extracts the temperature data of the bilayer structure's surroundings, provided to the input layer $\boldsymbol{T}^{(0)}$ of ANN. When reading the input data $\boldsymbol{T}^{(0)}$, the pre-trained ANN program in the computing system calculates the rotating velocity $\omega_1$ of PDMS in the core region $\Omega_1$ of the bilayer structure. The corresponding signal of controlling $\omega_1$ is transmitted to the stepper motor via the motor driver. Finally, the PDMS rotates around the center of the bilayer structure, driven by the stepper motor, thereby regulating the effective thermal conductivity of $\Omega_1$.

**Supporting Information**

Supporting Information is available from the Wiley Online Library or from the author.

**Acknowledgements**

J.H. acknowledges financial support from the National Natural Science Foundation of China under Grant No. 12035004, from the Science and Technology Commission of Shanghai Municipality under Grant No. 20JC1414700, and from the Innovation Program of Shanghai Municipal Education Commission under Grant No. 2023ZKZD06. C.W.Q. acknowledges financial support from the Ministry of Education, Republic of Singapore, under Grant No. A-8000107-01-00.

**Conflict of Interest**

The authors declare no conflict of interest.